\documentclass[a4paper,fleqn,usenatbib]{mnras}

\usepackage{newtxtext,newtxmath}
\usepackage[T1]{fontenc}
\usepackage{ae,aecompl}

\usepackage{graphicx}	% Including figure files
\usepackage{amsmath}	% Advanced maths commands
\usepackage{cleveref}
\usepackage{subfig}
\usepackage[utf8]{inputenc}		% accented letters
\usepackage{pifont}
\usepackage{romannum}
\usepackage[binary-units,range-units=single]{siunitx}
\DeclareSIUnit\parsec{pc}
\title[Lopsidedness in Early-Type Galaxies]{Lopsidedness in Early-Type Galaxies: the role of the $m=1$ multipole \\in Isophote Fitting and Strong Lens Modelling}

\author[A. Amvrosiadis et al.]{A. Amvrosiadis$^{1}$\thanks{E-mail: aristeidis.amvrosiadis@durham.ac.uk},
J. W. Nightingale$^{1, 2, 3}$,
Q. He$^{1}$, 
A. Robertson$^{4}$,
S. Lange$^{1}$,
C. S. Frenk$^{1}$, 
S. Cole$^{1}$, \newauthor
R. Massey$^{1, 2}$, 
A. Poci$^{5}$
\vspace{4mm}\\
$^{1}$ Institute for Computational Cosmology, Department of Physics, Durham University, South Road, Durham DH1 3LE, UK \\
$^{2}$ Centre for Extragalactic Astronomy, Department of Physics, Durham University, South Road, Durham DH1 3LE, UK \\
$^{3}$ School of Mathematics, Statistics and Physics, Newcastle University, Newcastle upon Tyne, NE1 7RU, UK \\
$^{4}$ Jet Propulsion Laboratory, California Institute of Technology, 4800 Oak Grove Drive, Pasadena, CA 91109, USA \\
$^{5}$ Sub-department of Astrophysics, Department of Physics, University of Oxford, Denys Wilkinson Building, Keble Road, Oxford, OX1 3RH, UK \\
}

\pubyear{???}

\begin{document}
\pagenumbering{arabic}
\label{firstpage}
%\pagerange{\pageref{firstpage}--\pageref{lastpage}}
\maketitle

% Abstract
\begin{abstract}
The surface brightness distribution of massive early-type galaxies (ETGs) often deviates from a perfectly elliptical shape. To capture these deviations in their isophotes during an ellipse fitting analysis, Fourier modes of order $m = 3, 4$ are often used. In such analyses the centre of each ellipse is treated as a free parameter, which may result in offsets from the centre of light, particularly for ellipses in the outer regions. This complexity is not currently accounted for in the mass models used in either strong gravitational lensing or galactic dynamical studies. In this work, we adopt a different approach, using the $m=1$ Fourier mode to account for this complexity while keeping the centres of all perturbed ellipses fixed, showing that it fits the data equally well. We applied our method to the distribution of light emission to a sample of ETGs from the MASSIVE survey and found that the majority have low $m_1$ amplitudes, below 2 percent. Five out of the 30 galaxies we analysed have high $m_1$ amplitudes, ranging from 2 to 10 percent in the outer parts ($R \gtrsim 3$ kpc), all of which have a physically associated companion. Based on our findings, we advocate the use of the $m=1$ multipole in the mass models used in strong lensing and dynamical studies, particularly for galaxies with recent or ongoing interactions.  
\end{abstract}


%%%%%%%%%%%%%%%%%%%%%%%%%%%%%%%%%%%%%%%%%%%%%%%%%%%%%%%%%%%%%%%%%%%%%%%%%%%
% SECTION
%%%%%%%%%%%%%%%%%%%%%%%%%%%%%%%%%%%%%%%%%%%%%%%%%%%%%%%%%%%%%%%%%%%%%%%%%%%
\section{Introduction} \label{sec:section_1}

The most massive early-type galaxies (ETGs) in the local Universe emerged from high density regions of the primordial matter distribution \citep{Blumenthal_84, Frenk_85}. Galaxy interactions and mergers, which are expected to be frequent in dense environments, affect their morphologies and kinematics.

ETGs are approximately elliptical in shape and are typically modelled as such. However, a large fraction of ETGs have surface brightness profiles that exhibit deviations from  perfect ellipses. These deviations are evident in the isophotes, which can take on boxy or disky shapes \citep[][]{1987A&A...177...71B, 1988A&AS...74..385B, 1989A&A...217...35B, 2006MNRAS.370.1339H, 2014ApJ...787..102C, 2018ApJ...856...11G}, as well as in variations of the position angle of their isophotes as a function of radius, known as isophotal twists \citep[][]{2002MNRAS.333..400C}. These azimuthal asymmetries likely reflect the complex gravitational dynamics at play during the formation and evolution of the galaxy and it is unsurprising that they often exhibit radial trends. 

One approach to parametrize these deviations is through the use of high order Fourier modes (i.e. multipoles; see Section~\ref{sec:section_3}), typically of order $m=3, 4$ \citep[e.g.][see also Multi-Gaussian Expanion; MGE]{2006MNRAS.370.1339H, 2018ApJ...856...11G}. For instance, the $m=4$ multipole can account for boxiness or diskiness in the light distribution \citep[e.g.][]{1987A&A...177...71B, 1988A&AS...74..385B, 1989A&A...217...35B}. The typical amplitudes of these multipoles are found to be up to $\sim$1 percent \citep[e.g.][]{2006MNRAS.370.1339H, 2006ApJ...636..115P, 2018ApJ...856...11G}. It is important to note that studies employing an ellipse fitting technique to analyze the isophotes of ETGs allow the centres  of the ellipses to vary. In a few cases, fractional offsets, $\Delta R / R$, are found, which can be as large as 10 to 15 percent at $R > 5$ kpc \citep[Figure 3 in ][]{2018ApJ...856...11G}. We will refer to this later type of complexity as lopsidedness.

The types of complexity discussed above are also expected to be present in the mass distribution of galaxies. Strong gravitational lensing is a powerful tool for constraining the mass distribution of galaxies acting as lenses. Utilizing strong lensing, many recent studies have shown that angular complexity in the mass distribution, in the form of $m=3,4$ multipoles, is strongly favoured by the data \citep[e.g.][]{2016ApJ...823...37H, 2022A&A...659A.127V, 2022MNRAS.516.1808P, 2024MNRAS.52710480N, 2024MNRAS.528.7564B, 2024arXiv240304850S, 2024arXiv240308895C}, with the level of correction matching those found from ellipse fitting to the light distribution ($\sim 1-2$ percent). However, commonly used mass models in strong lensing lack the flexibility to account for features such as lopsidedness (isodensity contours are assumed to share a common centre), and therefore some of these measurements could be biased. The lens models used in \citet{Nightingale2019} and \cite{Nightingale2023} favoured stellar mass profiles with centre offsets of order $100 - 300$pc, which the authors suggested is evidence for lopsidedness in the mass distribution.

\begin{figure}
\includegraphics[width=0.950\columnwidth,height=0.275\textheight]{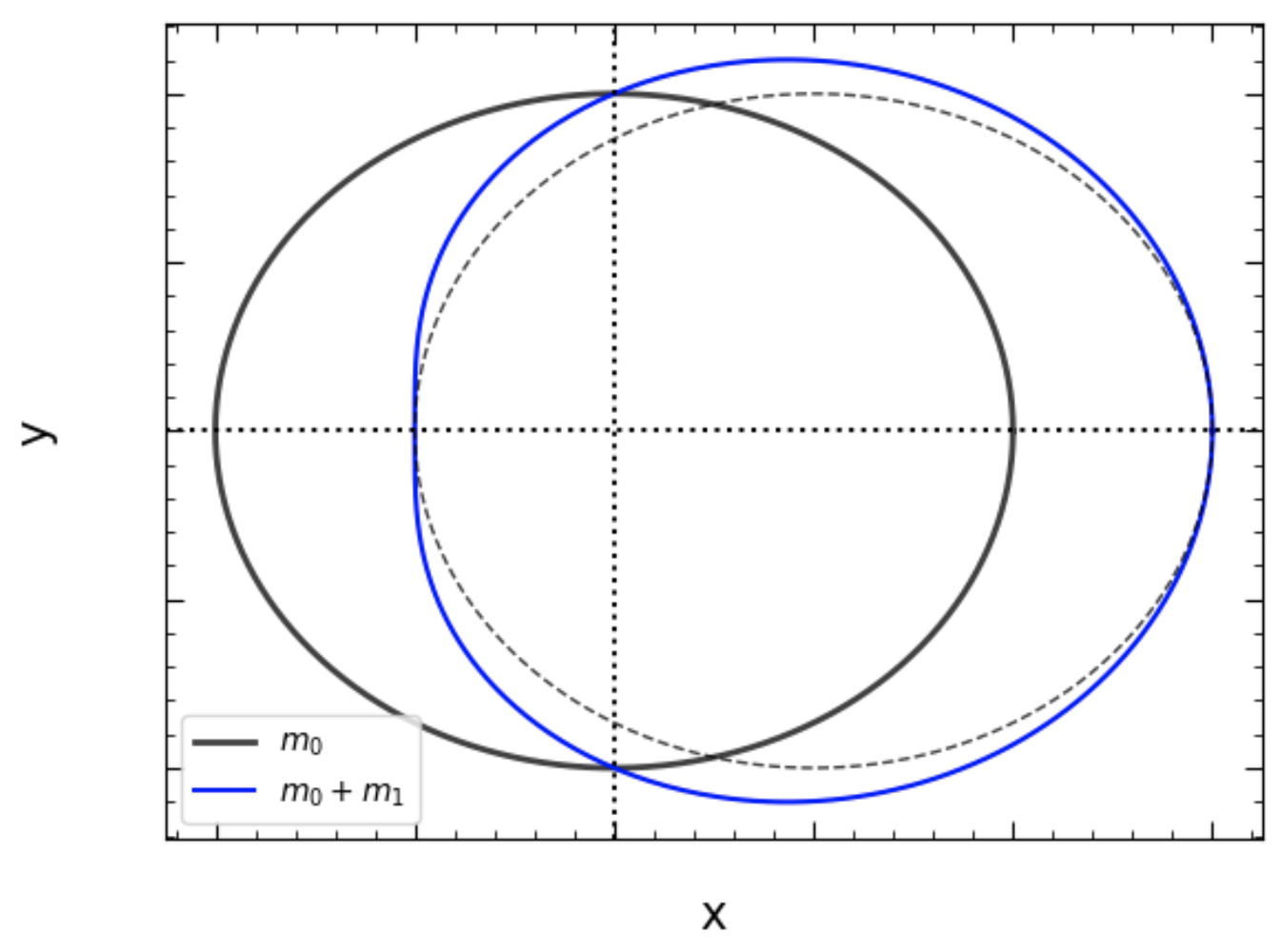}
\caption{Example of how a circular path, $m_0$ (black; solid), is perturbed by an $m=1$ multipole (blue; solid), where the position angle, $\phi_1$, of the $m_1$ mode is aligned with the x-axis. For comparison, we show a circular path that has been shifted in the x-axis (black; dashed) by the same amount as the $m_1$ amplitude.}
\label{fig:fig_0}
\end{figure}

In this work, we propose the use of the $m=1$ multipole to account for lopsidedness in the light/mass distribution of ETGs. Our objective is to showcase the utility of the $m=1$ multipole in ellipse fitting analyses as an alternative to allowing the centres of ellipses to vary (the former can be easily implemented in mass models used for strong gravitational lensing). To achieve this, we conduct an isophotal analysis of a sample of the most massive ETGs in the local Universe from the MASSIVE survey. It is important to note that an $m=1$ Fourier mode is not equivalent to a positional shift of the ellipse's centre (as demonstrated in Fig.~\ref{fig:fig_0}). We expect these to not be fully independent, especially in strong lensing studies (Lange et al. in prep.; Amvrosiadis et al. in prep.).

The outline of the paper is as follows. In Section~\ref{sec:section_2} we introduce the sample used in this work. In Section~\ref{sec:section_3} we describe our method to extract isophotal fit parameters. In Section~\ref{sec:section_4} we present our results. Finally, in Section~\ref{sec:section_5} we discuss these results and place them in the context of strong lensing and galactic dynamics. Finally, in Section~\ref{sec:section_6} we present our conclusions.  Throughout this work, we adopt a spatially-flat $\Lambda$-CDM cosmology with H$_0=67.8 \pm 0.9$\,km s$^{-1}$ Mpc$^{-1}$ and $\Omega_{\rm M}=0.308 \pm 0.012$ \citep{2016A&A...594A..13P}.

%%%%%%%%%%%%%%%%%%%%%%%%%%%%%%%%%%%%%%%%%%%%%%%%%%%%%%%%%%%%%%%%%%%%%%%%%%%
% SECTION
%%%%%%%%%%%%%%%%%%%%%%%%%%%%%%%%%%%%%%%%%%%%%%%%%%%%%%%%%%%%%%%%%%%%%%%%%%%
\section{Observations} \label{sec:section_2}

The sample we adopt was first introduced in \cite{2018ApJ...856...11G}, which is comprised of 35 galaxies selected from the MASSIVE survey \citep[][]{2014ApJ...795..158M}, a volume-limited ($D < $\,108\,Mpc) survey of 116 early-type galaxies. The galaxies in this sample were observed with the Hubble Space Telescope (HST), for one orbit, in the F110W filter with exposure times between $2500$ to $2900$~sec.

We excluded a few sources in our analysis from the original sample of \cite{2018ApJ...856...11G}: NGC0545, NGC0547, NGC1684, NGC5353, and NGC6482. These exclusions were due to our inability to obtain a good fit to their isophotes. However, it is important to note that the primary aim of this work is to demonstrate the effectiveness of the $m_1$ multipole compared to a variable ellipse centre for isophotal analysis (as discussed later in Section~\ref{sec:section_4}). Therefore, removing sources that are challenging to fit, regardless of whether using an $m_1$ or variable centre, does not diminish the validity of our overall conclusions.

%%%%%%%%%%%%%%%%%%%%%%%%%%%%%%%%%%%%%%%%%%%%%%%%%%%%%%%%%%%%%%%%%%%%%%%%%%%
% SECTION
%%%%%%%%%%%%%%%%%%%%%%%%%%%%%%%%%%%%%%%%%%%%%%%%%%%%%%%%%%%%%%%%%%%%%%%%%%%
\section{Isophote fitting} \label{sec:section_3}

% NOTE: 
\begin{figure}
\begin{tabular}{c}
\includegraphics[width=0.475\textwidth,height=0.165\textheight]{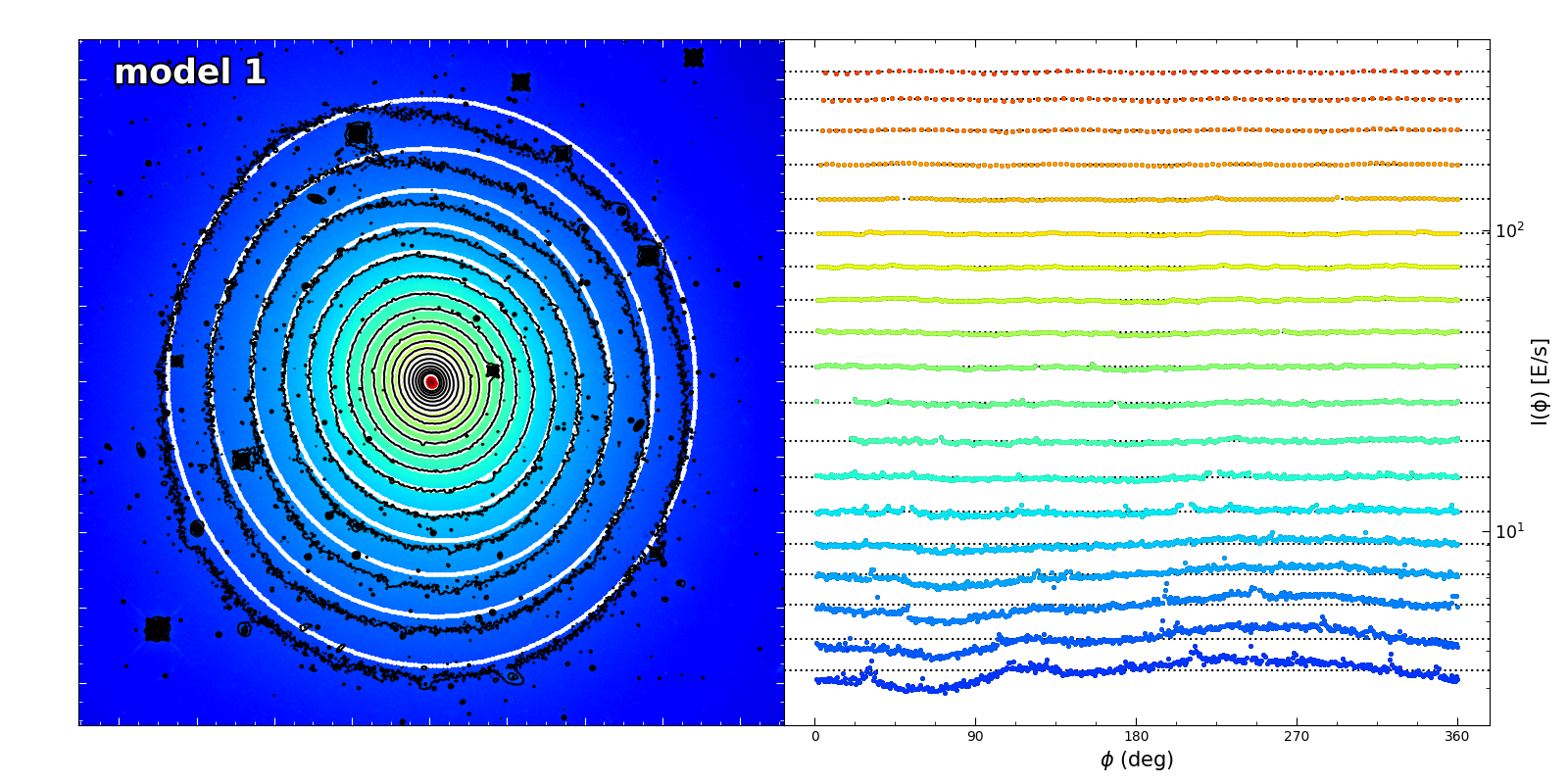} \\[-4mm]
\includegraphics[width=0.475\textwidth,height=0.165\textheight]{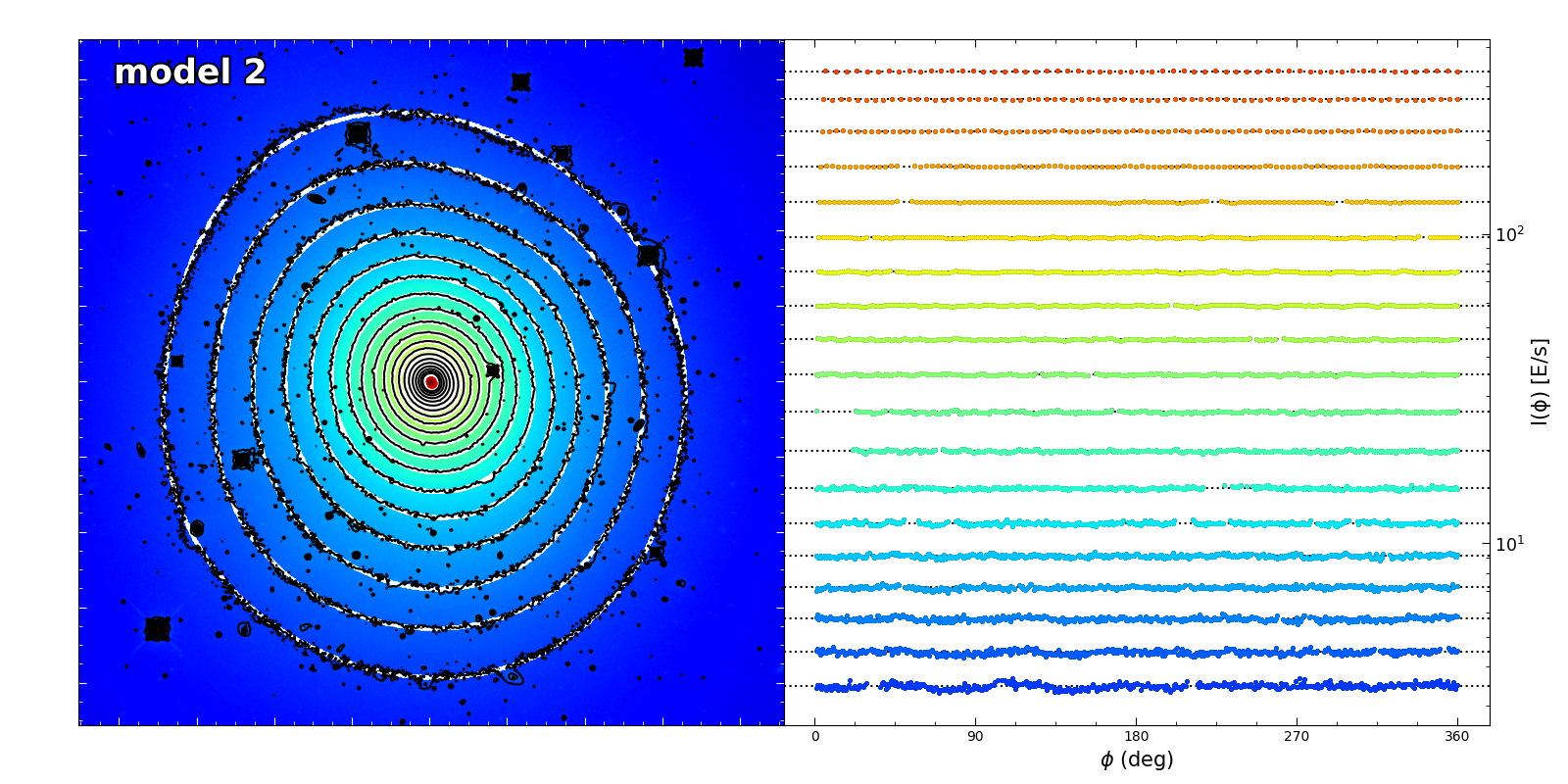} \\[-4mm]
\includegraphics[width=0.475\textwidth,height=0.165\textheight]{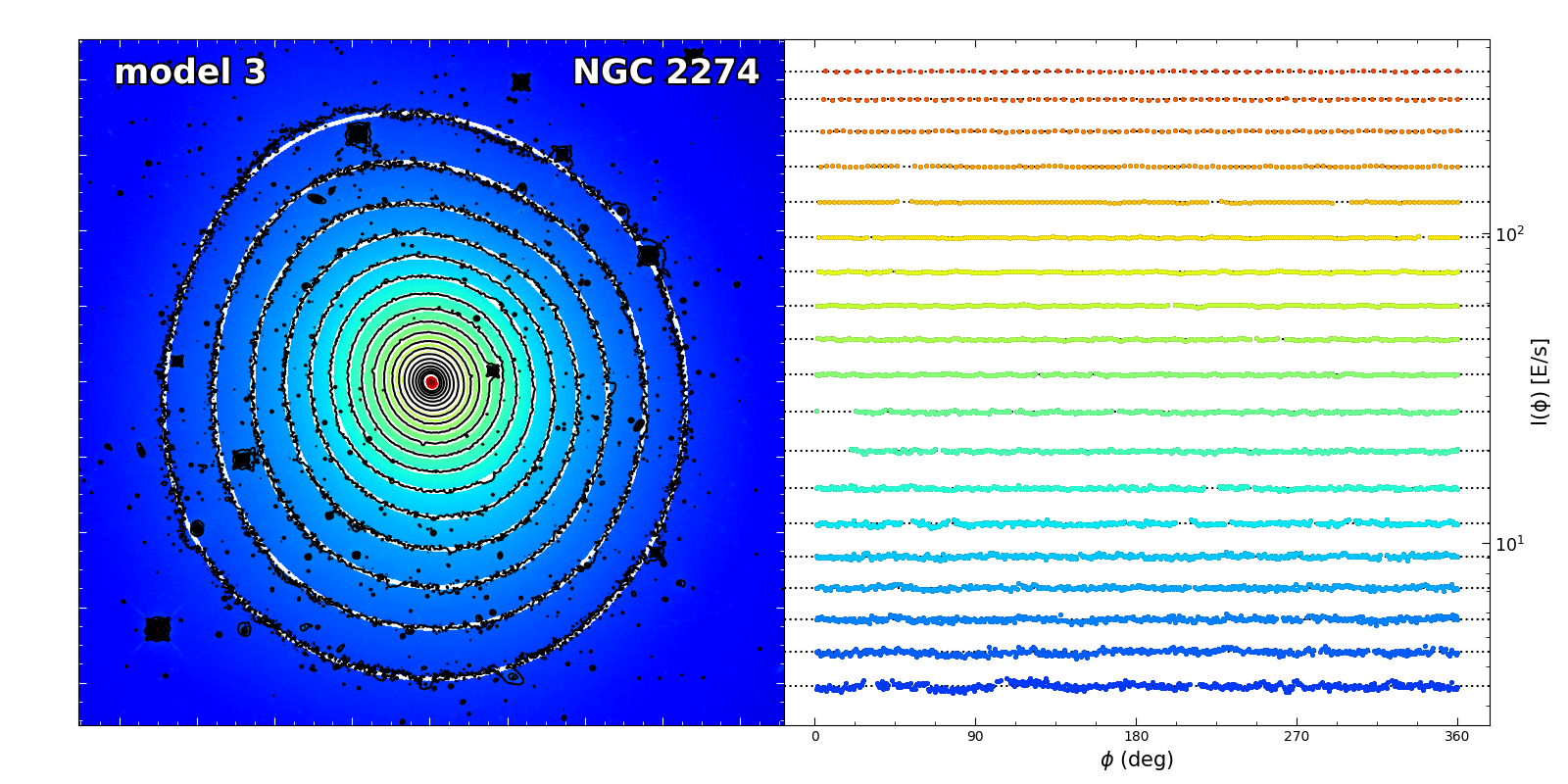} \\
\includegraphics[width=0.995\columnwidth,height=0.225\textheight]{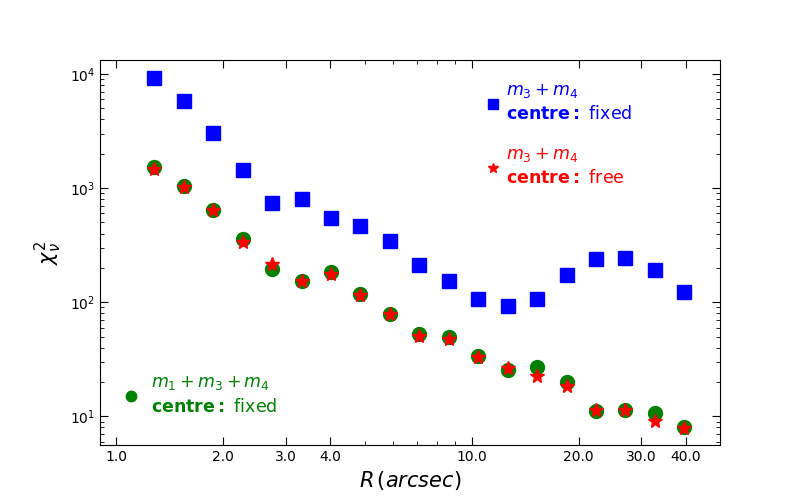} \\
\end{tabular}
\caption{Example of our ellipse fitting method applied to NGC2270. The panels from top to bottom correspond to models: (i) multipoles: $m=3,4$; centre: fixed, (ii) multipoles: $m=3,4$; centre: free and (iii) multipoles: $m=1,3,4$; centre: fixed. The bottom panel show the reduced chi-square of isophotes for the above three models, as a function of galactocentric radius.}
\label{fig:ellipse_fitting_example}
\end{figure}

\begin{figure*}
\centering
\begin{tabular}{c}
\hspace{0.525cm}\includegraphics[width=0.85\textwidth,height=0.13\textheight]{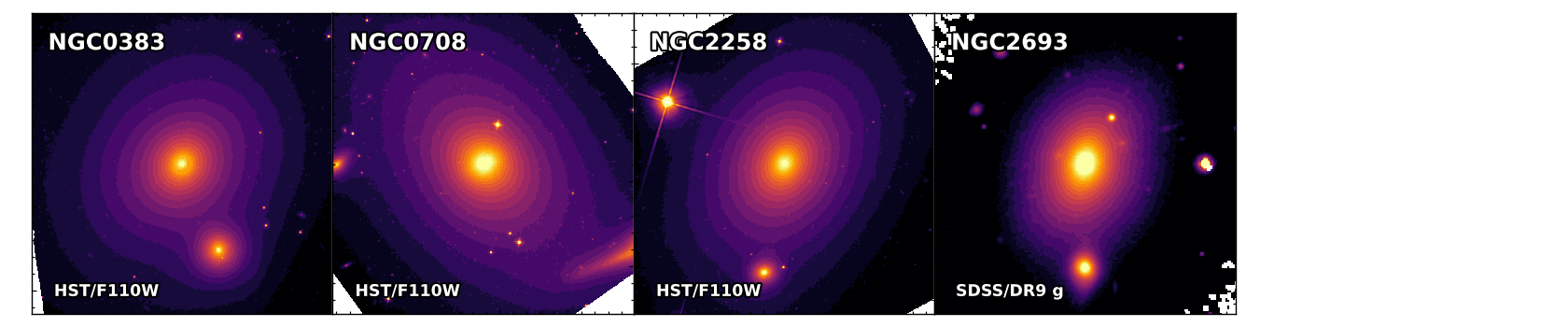} \\[-2.25mm]
\hspace{0.525cm}\includegraphics[width=0.85\textwidth,height=0.13\textheight]{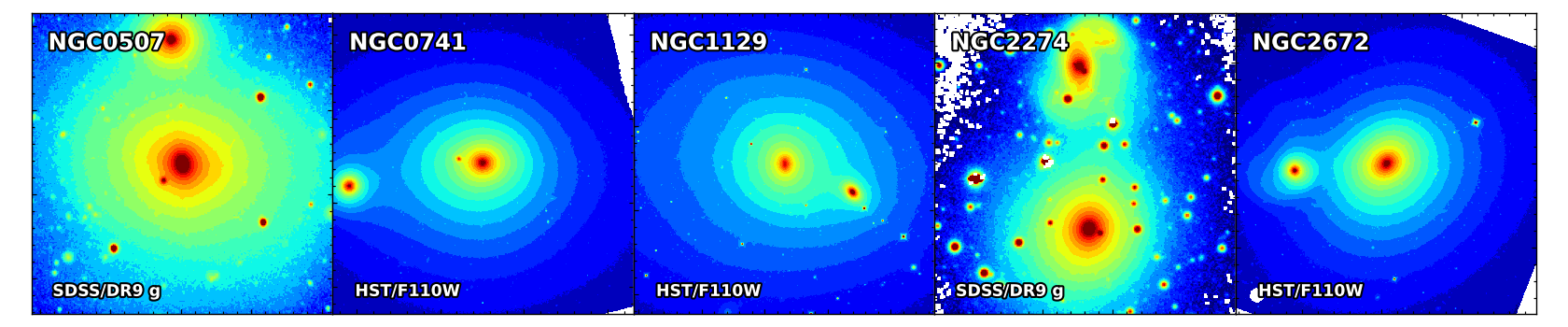} \\[-1.5mm]
\includegraphics[width=0.925\textwidth,height=0.36\textheight]{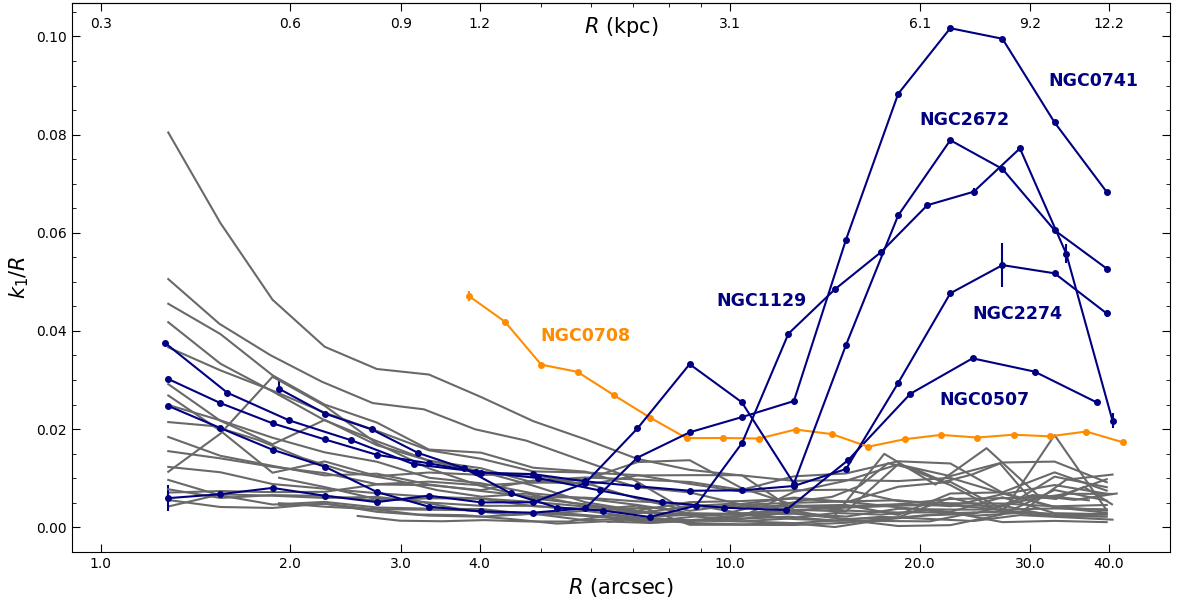} \\
\end{tabular}
\caption{The $m_1$ amplitude radial profiles for all sources in our sample. We show in blue galaxies with a visible companion galaxy, encompassing all those with high $(>2\%)$ $m_1$ amplitudes. NGC0708, shown in orange, has the most pronounced dust lanes. The top row cutouts show galaxies with companions and low $m_1$ amplitude, while the second row show galaxies with high $m_1$ amplitudes. For some of these galaxies the field of view of HST is too small to see the companion (e.g. \ NGC 2274). In these cases we show a g band image taken from the legacy surveys. North in up in all images.}
\label{fig:m1}
\end{figure*}

We developed our own routine to perform the ellipse fitting analysis which simultaneously optimizes the parameters of the ellipse ($\boldsymbol{\theta}_{\rm e}$: $x_0$, $y_0$, $e$, $\theta$), and the parameters of the multipole perturbations ($\boldsymbol{\theta}_{\rm m}$: $a_m$, $b_m$ for $m=1,3,4$) and builds on previous methods \citep[e.g.][]{1987A&A...177...71B, 2006MNRAS.370.1339H, 2018ApJ...856...11G}. 

The equation of an ellipse in polar coordinates, $r_{\rm e}(\phi)$, is given by,
\begin{equation}
    r_{\rm e}(\phi) = \frac{a (1 - e^2)}{\sqrt{1 + e \cos\left(\phi - \theta\right)}} \, ,
\end{equation}
where $a$ is the major axis, $e$ is the ellipticity and $\theta$ is the position angle of the major axis. We add perturbations to that ellipse in the form of multipoles given by,
\begin{equation} 
    r_m(\phi) = a_m \cos\left[m \left(\phi - \theta \right)\right] + b_m \sin\left[m \left(\phi - \theta \right)\right] \, ,
\end{equation}
where $m$ is the harmonic mode. We only consider multipoles of order $m = 1, 3, 4$ as the $m=2$ is already included in the model (i.e. ellipticity). 

From the above equations, the perturbed elliptical path in polar coordinates is given by, $r(\phi) = r_{\rm e}(\phi) + r_m(\phi)$. In a Cartesian system, the $(x, y)$ coordinates of the perturbed elliptical path can then be computed as, 
\begin{align*} 
    x = x_0 + r(\phi) \cos\phi \\
    y = y_0 + r(\phi) \sin\phi
\end{align*}
where ($x_0$, $y_0$) correspond to the centre of the ellipse. 

We optimize the parameters of the perturbed ellipse for different fixed values of the major axis, $a$, which are logarithmically spaced. During optimization we aim to maximize the log of the likelihood function given by,
\begin{equation}
    L = -\sum_{\phi = 0}^{360}\left(\frac{\rm{I}(\phi | \boldsymbol{\theta}_{\rm e}, \boldsymbol{\theta}_{\rm m}) - \langle \rm{I}(\phi | \boldsymbol{\theta}_{\rm e}, \boldsymbol{\theta}_{\rm m}) \rangle}{\sigma_{\rm{I}}}\right)^2 \,
\end{equation}
where $\rm{I}(\phi | \boldsymbol{\theta}_{\rm e}, \boldsymbol{\theta}_{\rm m})$ is the intensity profile along the perturbed elliptical path constructed from the $(\boldsymbol{\theta}_{\rm e}, \boldsymbol{\theta}_{\rm m})$ set of parameters\footnote{The observed images are first interpolated on a regular grid so that we can sample the intensity at any (x, y) location.} with $\langle\rm{I}\rangle$ representing the median of these values\footnote{Subtracting the mean of values along the path reduces the number of parameters we fit for and is equivalent to requiring that the intensity is constant along the path (the definition of an isophote). Alternatively, one can introduce an additional parameter for the intensity as it is typically done.} and $\sigma_{\rm{I}}$ is the error map (we use the WHT map of the HST data products). 

The parameter optimization is carried out using the {\it emcee} package \citep[][]{2013PASP..125..306F} for each of the isophotes independently. The centre of the perturbed ellipses is kept fixed during optimization to the value we estimate for the innermost ellipse. A few galaxies in this sample have dust lanes in their central regions, which can prevent an accurate centre determination. In these cases the innermost ellipse is shifted outwards. 

%%%%%%%%%%%%%%%%%%%%%%%%%%%%%%%%%%%%%%%%%%%%%%%%%%%%%%%%%%%%%%%%%%%%%%%%%%%
% SECTION
%%%%%%%%%%%%%%%%%%%%%%%%%%%%%%%%%%%%%%%%%%%%%%%%%%%%%%%%%%%%%%%%%%%%%%%%%%%
\section{Results} \label{sec:section_4}

\begin{figure*}
\centering
\includegraphics[width=0.95\textwidth,height=0.16\textheight]{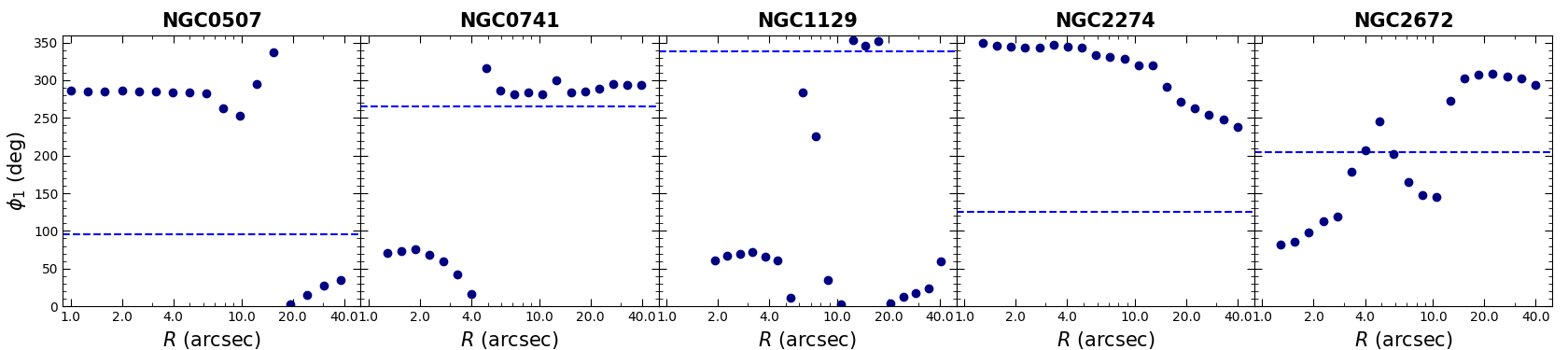}
\caption{The position angle of the $m_1$ multipole as a function of radius for the five galaxies with the largest $m_1$ amplitudes. The dashed horizontal lines correspond to the position angle of the companion galaxy.}
\label{fig:m1_angles}
\end{figure*}

% \begin{table}
% %\renewcommand{\arraystretch}{1.5}
% \centering
% \begin{tabular}{l|c|c}

%  & modes & centre fixed  \\
% \hline

% \textbf{model 1} & 3, 4 & \xmark \\
% \textbf{model 2} & 3, 4 & \cmark \\
% \textbf{model 3} & 1, 3, 4 & \cmark \\
% \end{tabular}
% \caption{}
% \label{tab:models}
% \end{table}

In Figure~\ref{fig:ellipse_fitting_example} we show an example of our isophotal analysis applied to NGC2274. We examine 3 cases: \textbf{(model 1)} $m=3,\ 4$ and fixed centre, \textbf{(model 2)} $m=3,\ 4$ and free centre, \textbf{(model 3)} $m=1,\ 3,\ 4$ and fixed centre. For model 1, the extracted isophotal intensity profile notably deviates from a flat trend, particularly evident for the outermost ellipses ($R > 10$ arcsec; approximately 3 kpc at $z = 0.015$; right-hand panel in the top figure), indicating a clear first-order perturbation in the extracted isophotes. In contrast, in model 2 where we allow the centre to vary freely, the fit improves significantly. We observe an increasing fractional offset for ellipses ($\Delta R / R$) as distance from the galactic centre increases, consistent with findings in \cite{2018ApJ...856...11G}. Finally, both model 2 and 3 provide equally satisfactory fits to the data, assessed using the reduced chi-square as the figure of merit, as illustrated in the bottom panel of Figure~\ref{fig:ellipse_fitting_example}. This suggests that the $m=1$ parameterization can effectively describe the light (or mass) distribution of ETGs without necessitating a variable centre. Moreover, this parameterization offers a straightforward application for mass models used in strong lensing and galaxy dynamics, which can be physically interpreted as lopsidedness (or skewness).

In Figure~\ref{fig:m1}, we present the radial profiles of $m_1$ amplitudes extracted from our isophotal analysis for all the galaxies we analyzed. The overall shape of the $m_1$ amplitude profiles displays similar features to the fractional offset profiles in \cite{2018ApJ...856...11G}. The peak $m_1$ amplitude we measure is $\sim$10 percent, which is slightly lower compared to the maximum fractional offset, ${\rm max}\{\Delta R / R\} \sim$15 percent, in \cite{2018ApJ...856...11G}. The majority of these profiles exhibit a radial decline in the inner regions extending out to $R\sim5$ arcsec, with the rest remaining flat (this trend is also seen in Goullaud et al. 2018). Beyond $R > 5$ arcsec, the $m_1$ profiles remain flat, for most galaxies, with amplitudes below 2 percent. Among the 30 sources analyzed, only five display a gradual increase in $m_1$ amplitude with radius, peaking around $20 - 30$ arcsec ($\sim$6$-$10 kpc for $z=0.015$). In the following section we discuss the origin for the high $m_1$ amplitudes in these galaxies. 

Although we do not show the profiles for the $m=3$ and $m=4$ multipoles, we note that these are consistent with those in \cite{2018ApJ...856...11G}. Therefore, irrespective of how we parametrize the isophotal path (i.e.\ either as a radial offset or an $m=1$ multipole), higher order multipole measurements remain unchanged.

%%%%%%%%%%%%%%%%%%%%%%%%%%%%%%%%%%%%%%%%%%%%%%%%%%%%%%%%%%%%%%%%%%%%%%%%%%%
% SECTION
%%%%%%%%%%%%%%%%%%%%%%%%%%%%%%%%%%%%%%%%%%%%%%%%%%%%%%%%%%%%%%%%%%%%%%%%%%%
\section{Discussion} \label{sec:section_5}

In this section we discuss the possible physical origin of an $m=1$ multipole and the possible consequences of not including this type of complexity in mass models used in strong gravitational lensing and galaxy dynamics. 

%%%%%%%%%%%%
% subsection
%%%%%%%%%%%%
\subsection{The origin of an $m = 1$ multipole}

In the top panels of Figure~\ref{fig:m1} we show cut-outs of some of the galaxies in our sample. The top row shows galaxies with companions and low $m_1$ amplitudes (<2\%), while the bottom row shows all galaxies with high $m_1$ amplitudes. These all have a physically associated companion $\sim$10$-$40 kpc away from their galactic centre in projection. This suggests that high $m_1$ amplitudes (or ellipse offsets) are the result of tidal interactions, where the outermost stellar orbits are highly eccentric \citep{2018ApJ...856...11G}. Using K-band magnitudes, $\rm M_{\rm K}$, we estimate the stellar mass ratios (between the main galaxy and its companion), for those with high $m_1$ amplitudes, to lie in the range $\sim 10 - 250$ (see Table~\ref{tab:sample}). 

We also tested whether we are artificially measuring high $m_1$ amplitudes due to light contamination from companions. To conduct this test, we created simulations of physically unassociated galaxies (i.e., superposition of two galaxies) with the same signal-to-noise ratio as the real data. Subsequently, we applied our ellipse fitting analysis to the main galaxy and found an $m_1$ amplitude profile consistent with zero. This test confirms that the physical origin of $m=1$ perturbations is galaxy interactions (recent or ongoing).

In Figure~\ref{fig:m1_angles} we show the $m_1$ position angle, $\phi_1$, profiles for all galaxies with high $m_1$ amplitudes. We find sharp transitions between the inner and outer regions, suggesting that physical processes (e.g. mergers, tidal interactions), occurring at different epochs, are shaping the profiles of ETGs. The $m_1$ position angle in the outer parts are often offset relative to the direction of the companion galaxy. This is perhaps not surprising when considering the relative timescales of stellar orbits and the motion of the companion. 

%%%%%%%%%%%%
% subsection
%%%%%%%%%%%%
\subsection{Strong Gravitational Lensing}

In galaxy scale strong gravitational lensing studies, the most common type of profile assumed is an elliptical power-law model \citep[EPL;][]{2015A&A...580A..79T}. Recently, a few studies have shown that assuming an EPL model for the mass distribution of the lens might not be enough to fit the data \citep[e.g.][]{2024MNRAS.528.7564B, 2024MNRAS.52710480N, 2024arXiv240304850S}; this becomes increasingly the case with higher resolution data \citep[e.g.][]{2022MNRAS.516.1808P, 2023MNRAS.518..220H}. As a result, attention has turned to the multipole model, particularly of order $m=3,4$, as a means for  accounting for additional complexity (e.g. angular deviations) in the mass distribution.

\begin{figure}
\centering
\includegraphics[width=0.995\columnwidth,height=0.2\textheight]{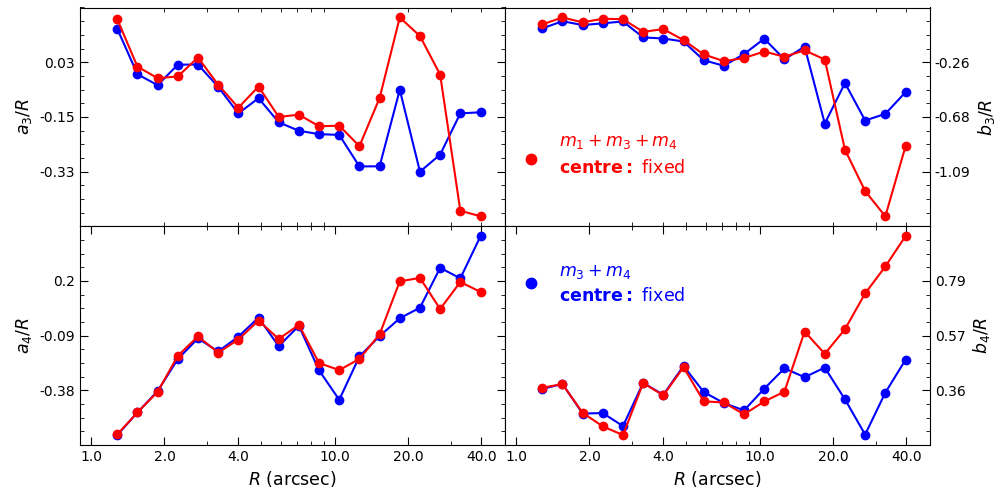}
\caption{Radial profiles of the Fourier coefficients, $a_m$ and $b_m$, of order $m=3,\ 4$, for models 1 (blue) and 3 (red) measured for NGC2274. If the $m_1$ mode is not taken into account, the measurements of higher order multipoles are biased in the outer parts.}
\label{fig:model_comparison}
\end{figure}

Models incorporating $m=3,\ 4$ multipoles define isodensity contours with respect to a common centre, placing them in the model 1 category. However, as demonstrated in Section~\ref{sec:section_4}, this model fails to capture observed features such as lopsidedness in the outer parts of galaxies, where model 3 (which includes an $m=1$ multipole) performs significantly better. \cite{2024arXiv240304850S} showed that $m=3,\ 4$ multipoles in the mass distribution of three strongly lensed galaxies are favoured by the data (according to the Bayesian evidence). However, the multipole's Fourier coefficients (or amplitude and position angles) inferred from an isophotal analysis are not consistent with the strong lensing results. The isophotes in these galaxies display large radial offsets that are not captured in strong lens modelling. As an example, we show in Figure~\ref{fig:model_comparison} that the $m=3,\ 4$ radial profiles of the Fourier coefficients differ in NGC2274 for model 1 \& 3, which would correspond to strong lensing and isophotal analyses, respectively. This missing complexity in the mass distribution might explain the differences seen in \cite{2024arXiv240304850S}.

An important consequence of ignoring the $m_1$ multipole complexity in mass models was recently explored by \cite{2024MNRAS.528.1757O}. These authors simulate strong lenses using a model that includes an $m=1$ perturbation, then fit the simulated data using a model that does not include $m_1$ but instead includes a dark matter subhalo, parametrized by a Navarro-Frenk-White profile \citep[NFW;][]{1996ApJ...462..563N, 1997ApJ...490..493N}. The motivation behind this analysis was to quantify the rate of "false-positive" detections of dark matter subhalos that can occur if the mass model for the lensing galaxy lacks complexity. Their findings suggest that the probability of false positives for lenses with $\sim$10 percent $m_1$ amplitudes is almost an order of magnitude higher than for models with $\sim$1 percent $m_4$ amplitudes (which are the typical values). It is therefore important that both these types of complexity be considered when modelling the lensing galaxy.

For galaxies with high $m_1$ amplitudes, the profiles typically peak around $\sim$6$-$10 kpc. For a galaxy-scale lens at $z=0.2/0.8$, an Einstein radius of $\rm r_E\sim1$ arcsecond corresponds to $\rm r_E \sim 3.4/7.7$ kpc in physical units. Given that strong lensing is particularly sensitive to the mass distribution near the Einstein radius, $m_1$ amplitudes of up to $\sim$10~percent are to be expected when modeling strong lenses.

%%%%%%%%%%%%
% subsection
%%%%%%%%%%%%
\subsection{Galactic Dynamics} 

The presence of an \(m=1\) multipole in real galaxies could also affect  dynamical models.  This is true for both the technical aspects of generating the model, as well as the physical interpretation of the dynamics. Many dynamical models are restricted to axisymmetry \citep[e.g.\ Jeans anisotropic models;][]{2008MNRAS.390...71C} in the mass distribution, which implies a common centre for all mass components, no isophotal twists and certainly no \(m=1\) multipole (asymmetry). At the most massive end, real galaxies often exhibit isophotal twists, as well as misalignments between kinematic and photometric position angles, which are both evidence for triaxiality. Triaxial implementations of dynamical models \citep[e.g.][]{2022ApJ...926...30Q, 2022ApJ...928..178P} are able to account for such features. Even so, twists and misalignments are point-symmetric perturbations, meaning there is symmetry about the origin. Real asymmetry in the mass distribution resulting in a strong \(m=1\) multipole implies that the galaxy is out of dynamical equilibrium, since such a configuration is not stable under self-gravity. Unlike gravitational lensing, dynamical equilibrium is a fundamental assumption of any dynamical model and thus large \(m=1\) multipoles pose conceptual challenges to modelling the dynamics of such galaxies.

Modern observations of stellar kinematics (e.g.\ the MASSIVE sample) routinely reach the physical scales which here are found to exhibit large \(m=1\) modes. However, dynamical modelling of massive ETGs out to \(\sim \SI{10}{\kilo\parsec}\) has yet to be conducted systematically. One example was presented in \cite{2022ApJ...928..178P}, where IFU data for NGC~2693 was modelled by a triaxial orbit-superposition code. Interestingly, isophotal twists are measured and attributed to the impact of a companion galaxy. The magnitude of this effect on measurements of the dynamical mass and internal structure in the presence of \(m=1\) multipoles remains to be quantified. Given that they are most prevalent in the outermost luminous regions where the stellar density is low, it is likely that the impact on existing dynamical models is minimal. This is true for the vast majority of dynamical studies, which make measurements at or within the effective radius, \(R_{\rm e} \sim \SI{10}{\kilo\parsec}\). Conversely, dynamical constraints on dark matter would arise mostly from these outer low stellar density regions, and so those measurements might be more affected by unaccounted for asymmetries.

%%%%%%%%%%%%%%%%%%%%%%%%%%%%%%%%%%%%%%%%%%%%%%%%%%%%%%%%%%%%%%%%%%%%%%%%%%%
% SECTION
%%%%%%%%%%%%%%%%%%%%%%%%%%%%%%%%%%%%%%%%%%%%%%%%%%%%%%%%%%%%%%%%%%%%%%%%%%%

\section{Conclusions} \label{sec:section_6}

We have measured perturbations to the isophotes of ETGs in the local Universe, selected from the MASSIVE survey. A novel aspect of our analysis is the use of the $m=1$ multipole instead of allowing the centre of the elliptical isophotes to vary with distance from the galactic centre, as is common practice in ellipse fitting analyses \citep[e.g.][]{2006MNRAS.370.1339H, 2018ApJ...856...11G}. This parametrization enables the model to capture features such as lopsidedness/skewness, which are currently overlooked in mass models used in strong gravitational lensing and galactic dynamics studies.

We found that the amplitude of the $m_1$ multipole can reach values as high as 10 percent on scales relevant to strong lensing, but only in galaxies that are undergoing interactions with other galaxies. We argue that models aiming to describe the mass distribution of ETGs should incorporate this level of complexity. Failure to account for it can lead, for instance, to false-positive detections of dark matter subhalos  in strong lensing studies \citep[e.g.,][]{2024MNRAS.528.1757O} or to an inability for the model to  focus light rays on the source plane accurately (Nightingale et al. 2024 in prep.). In two upcoming works, Lange et al. (2024, in prep.) and Amvrosiadis et al. (2024, in prep.), we will fit mass models incorporating an $m=1$ order multipole perturbation to the distribution of mass in two strongly lensed dusty star-forming galaxies.

\begin{table*}
\centering
\begin{tabular}{lcccccccc}
%k_m_ext
name & z & $\rm M_{\rm K}$ & Companion & $\rm z_{\rm comp}$ & $\rm M_{\rm K, comp}$ & $\rm M_{\star} / M_{\star, \rm comp}$ & $\theta_{\rm proj}$ & $\rm D_{\rm proj}$ \\
& & (mag) & & & (mag) & &  (arcsec) & (kpc) \\
\hline
%NGC0057 & - & - & - & - & - & - & - & - \\
%NGC0315 & - & - & - & - & - & - & - & - \\
NGC0383$^*$ & 0.017005 & - & NGC0382 & 0.017442 & - & - & 34.20 & 12.4 \\ % 8.481 & - | -
%NGC0410 & - & - & - & - & - & - & - & - \\
\textbf{NGC0507} & 0.016458 & -26.04 & NGC0058 & 0.018433 & -24.86 & 15 & 87.61 & 32.1 \\ % 8.302 & 9.725 | 15, D = 8670
%NGC0533 & - &- & -& -& - & - & - & - \\
NGC0708$^*$ & 0.015886 & - & NGC0705 & 0.015057 & - & - & 66.37 & 21.6 \\
\textbf{NGC0741} & 0.018294 & - & NGC0742 & 0.019910 & - & - & 47.57 & 19.0 \\ % 8.295 & - | -, D = 7088
%NGC0777 & - & - & - & - & - & - & - & - \\
%NGC0890 & - & - & - & - & - & - & - & - \\
%NGC1016 & - & - & - & - & - & - & - & - \\
%NGC1060 & - & - & - & - & - & - & - & - \\
\textbf{NGC1129} & 0.017472 & -26.23 & VV 085 & 0.016568 & -23.82 & 255 & 25.98 & 9.3 \\ % 8.240 & 10.530 | 255, D = 3969
%NGC1167 & - & - & - & - & - & - & - & - \\
%NGC1272 & - & - & - & - & - & - & - & - \\
%NGC1453 & - & - & - & - & - & - & - & - \\
%NGC1573 & - & - & - & - & - & - & - & - \\
%NGC1600 & - & - & - & - & - & - & - & - \\
%NGC1700 & - & - & - & - & - & - & - & - \\
NGC2258$^*$ & 0.013539 & - & - & - & - & - & 39.14 & 11.2 \\ % 8.227 / - | -
\textbf{NGC2274} & 0.016785 & -25.70 & NGC2275 & 0.016435 & -24.59 & 13 & 115.51 & 40.3 \\ % 8.682 & 9.750 | 13, D = 1538
%NGC2513 & - & - & - & - & - & - & - & - \\
\textbf{NGC2672} & 0.014487 & - & NGC2673 & 0.012508 & - & - & 25.98 & 9.3 \\ %8.351 & -
NGC2693$^*$ & 0.016083 & -25.68 & NGC2694 & 0.016908 & -22.94 & 557 & 55.70 & 19.3 \\ % 8.604 / 11.460 | -, D =8703
%NGC4914 & - & - & - & - & - & - & - & - \\
%NGC5322 & - & - & - & - & - & - & - & - \\
%NGC5557 & - & - & - & - & - & - & - & - \\
%NGC6482 & - & - & - & - & - & - & - & - \\
%NGC7052 & - & - & - & - & - & - & - & - \\
%NGC7619 & - & - & - & - & - & - & - & - \\
\end{tabular}
\caption{(1) name (2) spectroscopic redshift (3) K band absolute magnitude computed as, $\rm M_{\rm K} = K - 5\log_{10}D - 25$, where K is taken from the 2MASS catalogue ("k\_m\_ext"), (4) name of the companion (5) spectroscopic redshift of the companion (6) K band absolute magnitude of the companion (7) stellar mass ratio using the relation, $\rm \log_{10} M_{\star} = 10.58 - 0.44 \left(M_{\rm K} + 23\right)$, to estimate stellar masses (Cappellari 2013) (8) angular distance on the sky between the main and companion galaxies (9) projected distance in kpc using the redshift of the main galaxy.}
\label{tab:sample}
\end{table*}

%%%%%%%%%%%%%%%%%%%%%%%%%%%%%%%%%%%%%%%%%%%%%%%%%%%%%%%%%%%%%%%%%%%%%%%%%%%
% SECTION
%%%%%%%%%%%%%%%%%%%%%%%%%%%%%%%%%%%%%%%%%%%%%%%%%%%%%%%%%%%%%%%%%%%%%%%%%%%
\section*{Acknowledgements}

AA, QH, CSF and SC are supported by ERC Advanced Investigator grant, DMIDAS [GA 786910] to C.S.\ Frenk.
We acknowledge support from STFC Consolidated Grant ST/X001075/1. This work used the DiRAC@Durham facility managed by the Institute for Computational Cosmology on behalf of the STFC DiRAC HPC Facility (www.dirac.ac.uk). The equipment was funded by BEIS capital funding via STFC capital grants ST/K00042X/1, ST/P002293/1, ST/R002371/1, and ST/S002502/1, Durham University and STFC operations grant ST/R000832/1. DiRAC is part of the UK National e-Infrastructure.

\section*{DATA AVAILABILITY}
All data used in this work are publicly available.

\section*{Software Citations}

This work uses the following software packages: \href{https://github.com/numpy/numpy}{{NumPy}}, \href{https://github.com/matplotlib/matplotlib}{{Matplotlib}}, \href{https://github.com/astropy/astropy}{{Astropy}}, \href{https://github.com/scipy/scipy}{{Scipy}},
\href{https://github.com/scipy/scipy}{{Aplpy}}.

%%%%%%%%%%%%%%%%%%%%%%%%%%%%%%%%%%%%%%%%%%%%%%%%%%%%%%%%%%%%%%%%%%%%%%%%%%%%%%
%%%%%%%%%%%%%%%%%%%%%%%%%%%%%%%%%%%%%%%%%%%%%%%%%%%%%%%%%%%%%%%%%%%%%%%%%%%%%%

\bibliographystyle{mnras}
\bibliography{main}

%%%%%%%%%%%%%%%%%%%%%%%%%%%%%%%%%%%%%%%%%%%%%%%%%%%%%%%%%%%%%%%%%%%%%%%%%%%%%%
%%%%%%%%%%%%%%%%%%%%%%%%%%%%%%%%%%%%%%%%%%%%%%%%%%%%%%%%%%%%%%%%%%%%%%%%%%%%%%
\appendix

\section{Higher order multipole profiles}

\begin{figure}
\centering
\begin{tabular}{c}
\includegraphics[width=0.925\columnwidth,height=0.2\textheight]{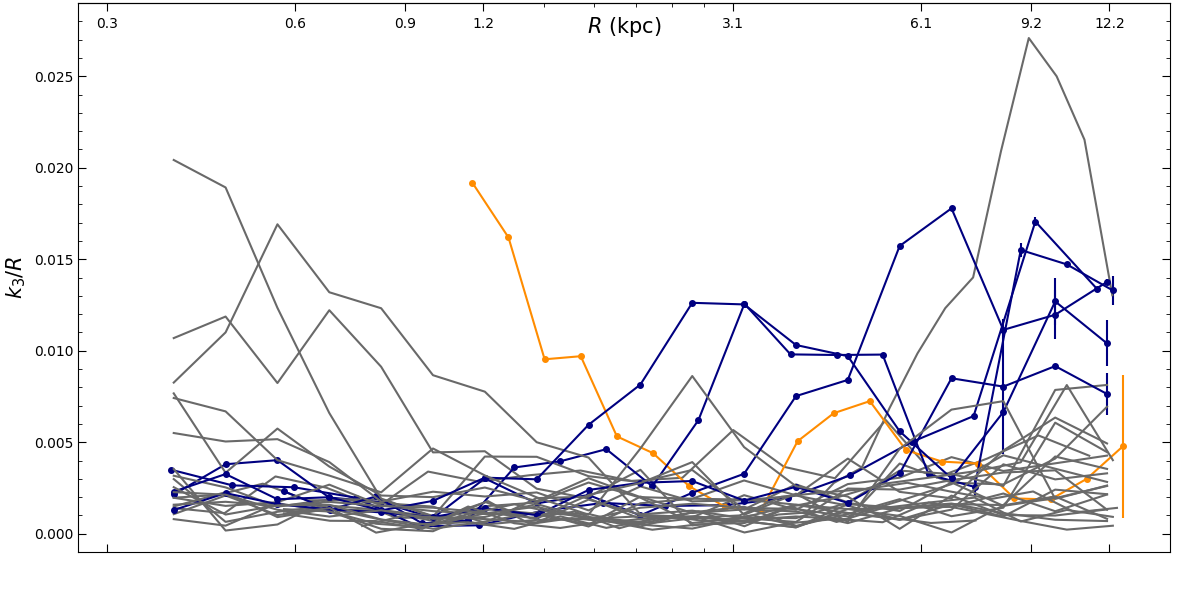} \\[-2.5mm]
\includegraphics[width=0.925\columnwidth,height=0.2\textheight]{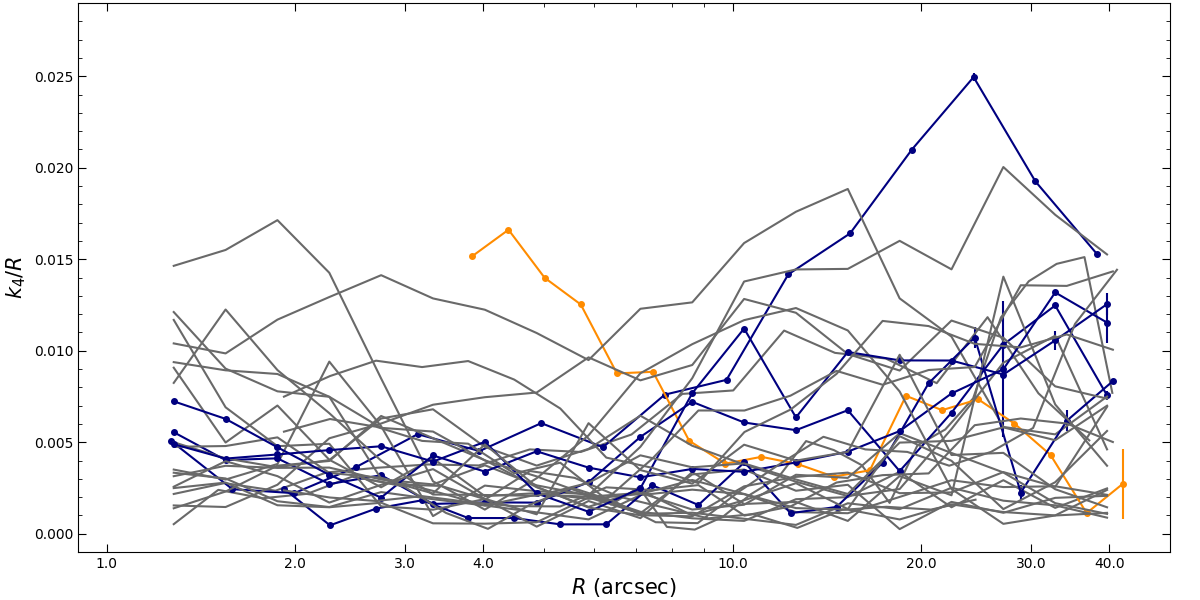} \\
\end{tabular}
\caption{Same as Fig.~\ref{fig:m1} but for the $m=3, \ 4$ multipoles. We colored in blue all galaxies with maximum $m=1$ amplitude above 2\% (bottom row cutouts in Fig.~\ref{fig:m1}).}
\label{fig:m34}
\end{figure}

In Figure~\ref{fig:m34} we show the radial profiles for the $m_3$ and $m_4$ multipoles. The maximum amplitudes of these higher order multipoles are consistent with previous studies \citep[$\sim$$2\%$; ][]{2018ApJ...856...11G}. In both panels, we colored in blue, galaxies that display elevated $m_1$ amplitudes (same as in Figure~\ref{fig:m1}). An interesting trend is that galaxies with high $m_1$ amplitudes also tend to exhibit the highest $m_3$ amplitudes at $R \gtrsim 3$ kpc\footnote{The galaxy with the highest maximum $m_3$ amplitude, NGC0383, has low $m_1$ amplitude. However, this galaxy also has a companion (see top panel in Fig~\ref{fig:m1}).}.This correlation, however, does not hold for the $m_4$ multipole.

This trend suggests that both $m=1$ and $m=3$ order multipole perturbations are short-lived and quickly dissipate, explaining the high amplitudes observed in galaxies with recent or ongoing interactions.  In contrast, $m=4$ order perturbations (e.g., bars in galaxies) are more stable and are therefore observed in more galaxies, regardless of recent or ongoing interactions with companions. 

%%%%%%%%%%%%%%%%%%%%%%%%%%%%%%%%%%%%%%%%%%%%%%%%%%%%%%%%%%%%%%%%%%%%%%%%%%%%%%
%%%%%%%%%%%%%%%%%%%%%%%%%%%%%%%%%%%%%%%%%%%%%%%%%%%%%%%%%%%%%%%%%%%%%%%%%%%%%%

\end{document}